# Improved circuits for a biologically-inspired random pulse computer

MARIO STIPČEVIĆ[1,*], MATEJA BATELIĆ[2]

[1]*Photonics and Quantum Optics, Center of Excellence for Advanced Materials and Sensing Devices, Ruđer Bošković Institute, Bijenička cesta 54, 10000 Zagreb, Croatia*
[2]*Department of Physics, Faculty of Science University of Zagreb, Bijenička cesta 32, 10000 Zagreb, Croatia*
**Corresponding author:* mario.stipcevic@irb.hr Date: 08/13/2019

***IMPORTANT NOTICE: This paper, with significant improvements, has been published in Scientific Reports and is available free of charge at this link*** **https://doi.org/10.1038/s41598-021-04177-9**
***Please cite the published work, as:*** Stipčević, M., Batelić, M. Entropy considerations in improved circuits for a biologically-inspired random pulse computer. Sci Rep **12,** 115 (2022). https://doi.org/10.1038/s41598-021-04177-9

**Abstract:** We present improved circuits intended for building a universal computer based on Random Pulse Computing (RPC) paradigm, a biologically-inspired (BIOLOGICALLY INSPIRED) way of computation in which variable is represented by a frequency of a Random Pulse Train (RPT) rather than a logic state. The RPC we mention here is also known as "stochastic unipolar computation" in newer literature. Unlike in previous art, where randomness is obtained from electronics noise or a pseudorandom shift register while processing circuitry is deterministic, in our approach both variable generation and signal processing rely on the random flip-flop (RFF) whose randomness is derived from a fundamentally random quantum process. This offers advantage in better precision and faster calculation.

## References and links

1. Solovay, Robert M.; Strassen, Volker, "A fast Monte-Carlo test for primality". SIAM Journal on Computing. **6,** 84–85 (1977) . DOI 10.1137/0206006.
2. Solovay, Robert M.; Strassen, Volker, "Erratum: A fast Monte-Carlo test for primality". SIAM Journal on Computing. **7**, 118 (1978). DOI 10.1137/0207009.
3. R. P. Feynman, "Simulating Physics with Computers", Int. J. Theor. Phys., **21**, 467-488 (1982). DOI 10.1007/BF02650179
4. J. von Neumann, "Probabilistic logics and the synthesis of reliable organisms from unreliable components". In Bródy, F.; Vámos, Tibor (eds.). *The Neumann Compendium*. World Scientific. pp. 567–616 (1995). ISBN 978-981-02-2201-7.
5. M. Stipčević, "Quantum random flip-flop and its applications in random frequency synthesis and true random number generation", Rev. Sci. Instrum. **87**, 035113 (2016). DOI 10.1063/1.4943668.
6. S. T. Ribeiro, "Random-Pulse Machines" IEEE Transactions on Electronic Computers, **16**, 261-276 (1967). DOI 10.1109/PGEC.1967.264662 .
7. B. R. Gaines, "Stochastic Computing Systems", Advances in Information Systems Science, **2**, 37-72 (1969) Ed. T. T. Julius, Springer US, Boston, MA. DOI 10.1007/978-1-4899-5841-9_2.
8. R. C Lawlor, "Computer utilizing random pulse trains", patent US3612845A, priority date Oct. 12, 1971.
9. A. Alaghi, J. P. Hayes, "Survey of stochastic computing", *ACM Trans. Embed. Comput. Syst.*, **12**, 92:1-92:19 (2013).
10. Y. Liu and K. K. Parhi. 2017. "Computing Polynomials Using Unipolar Stochastic Logic.", J. Emerg. Technol. Comput. Syst. **13**, 3, Article 42 (April 2017), 30 pages. DOI 10.1145/3007648.
11. B. Widrow, Y. Kim, D. Park, and J. Krause Perin, "Nature's Learning Rule: The Hebbian-LMS Algorithm", appears in Artificial Intelligence in the Age of Neural Networks and Brain Computing, Chaper 1, Pages 1-30, Eds.:Robert Kozma, Cesare Alippi, Yoonsuck Choe, Francesco Carlo Morabito, Academic Press, 2019. DOI 10.1016/B978-0-12-815480-9.00001-3.

## 1. Introduction

Today, the digital computing paradigm (DC), based on Turing machine theoretical model, is almost exclusively used. Implemented in electronics logic circuitry, that executes basic logic operations, and realized in solid-state chips, it allows very fast computation with an arbitrary precision. Since DC is incapable of generating randomness, a version enriched by a (single) random number generator, so called "randomized Turing machine", offers large speedup of certain tasks by use of randomized algorithms, for example testing primality of (big) numbers by Soloway-Strassen algorithm [1-2].

A radically new, quantum computation (QC) paradigm has been proposed by Feynman in 1981 [3]. It makes use of strong correlations of quantum entanglement and superposition principle to reach an exponential speed-up over DC of a small set of algorithms of a great practical importance. Input and output information to a QC is digital, however, internally it handles an analog construct: a multi-particle quantum state. The initial quantum state (the problem) is evolved very precisely by a set of operations to a final state (the solution) that is measured to obtain a statistical output. A large effort is being put on building a universal programmable quantum computer of precision that would have a practical significance, but thus far technological difficulties keep that goal out of reach.

The Random Pulse Computing (RPC) paradigm, proposed in a seminal work of John von Neumann [4] in 1956, shortly flashed in 1960s only to be run-over by the digital computation that flourished in 1970s after which RPC fell into oblivion. Re-born in mid 2010s, RPC can be thought of as a third computational paradigm. Input, output and internal information in the RPC are average frequencies of random pulse trains (RPT). The RPT is a sequence of logic pulses that appear randomly in time, wherein each pulse can be thought of as being a Poissonian random event. The only parameter describing RPT is its pulse rate. Even though it uses digital pulses and randomness, just as probabilistic DC, it radically differs from it. First, with a difference that information is not in form of a logic state of a register, but in the form of the RPT, the RPC fundamentally uses the time as a new dimension in calculation. Another, striking difference is that the RPC computer fundamentally and directly *computes* (makes mathematical operations like adding, squaring, etc.) while DC computer *does not compute*: a DC computer only makes a sequence of logic operations which may or may not represent a mathematical operation.

On top of that, RPC paradigm features a set of unique and peculiar characteristics that qualitatively resemble some of the characteristics of living beings, such as a need for a simple hardware for complex operations, immunity to errors in hardware components, massive parallelism and streamline computation (the output is formed continuously from the input), to name just a few. The main drive behind recent revival of the RPC is hope that it could efficiently (in terms of hardware and energy) solve problems that are currently difficult and/or energy consuming for the DC, but apparently easy for living beings, such as: learning, recognition and autonomous behavior. Conveniently, one now has a much better technological platform(s) for its realization than in 1960s. In our approach, along with the standard logic circuitry, quantum RFF [5] is a single additional circuit required to produce variables, random numbers, make random decisions and perform mathematical operations. By using a true, quantum randomness, the RPC may be seen as a bridge between DC and full-blown QC.

## 2. Random pulse train: probability versus frequency representation

In its original appearance [6-7], the RPT is a time-wise random sequence of logic pulses of constant height and width, as shown in Fig. 1a, wherein each positive-going edge corresponds to one Poissonian random event. Such events can be obtained from certain types of electronics noise passing a constant threshold [8], decay of a nuclei, etc. As mentioned before, a p.d.f. of waiting times is an exponential distribution with a single parameter: average frequency of pulses. As will be shortly discussed, the idea of RPC includes performing basic logic operations between two or more RPTs. This can be conveniently realized in field programmable gate array (FPGA) chips.

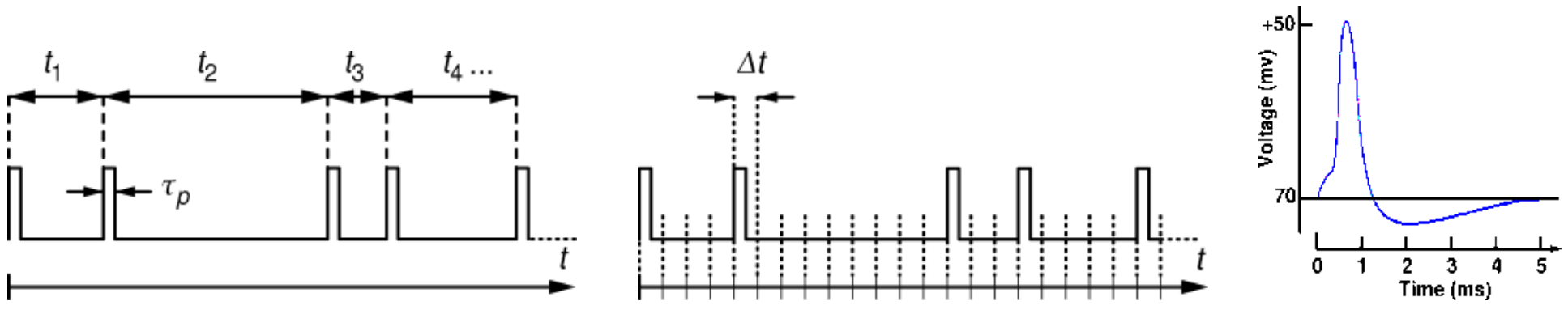

a) b) c)

**Fig. 1.** Random pulse train in which each pulse corresponds to a Poissonian random event with frequency $1/\langle t_i \rangle$ (a); Time-digitized RPT wherein appearance of a pulse in a given time bin of width $\Delta t$ is an independent binomial random event with probability $p$ (b); one full cycle of a biological nerve pulse in mammalians that lasts about 5 ms (c).

However, logic operations between two logic pulses that can appear at any time can easily result in output pulses of duration anywhere between zero and $2\tau_p$, or even pairs of short pulses, which presents a great technological inconvenience. In order to avoid this, a solution is found in digitizing the timeline in non-overlapping bins of duration $\Delta t$, as shown in Fig. 1b. Technically, a "clock" of frequency $f_{\mathrm{CLK}} = 1 \,/\, \Delta t$ is now defining time. Each time bin can hold one pulse. A probability $p \in [0,1]$ of a pulse to appear in a time bin, rather than frequency, is now the numerical information carried by the RPT. Note that pulses will still appear with a definite frequency, namely $p \,/\, \Delta t$, but time digitization ensures that pulses from various RPTs appear neatly aligned in time so that they can be unambiguously combined. That said, we should not forget that in biological systems there is no common clock that would align all impulses nor time digitization.

Even with time-digitized RPTs, a care has to be taken in order to avoid atomic-race conditions and appearance of spikes. Fortunately, such techniques are well known in the art of programming of FPGAs and therefore we will not address them in this paper.

In this work we only work with so-called unipolar representation of the RPT since it has a strongest resemblance to impulses found in mammalian nerve cells. Other representations known in literature are bipolar and stochastic single-line [7].

In the rest of this paper numbers are represented by probabilities of RPTs and are in the interval $[0,1]$. If different range of values is requited, some kind of mapping has to be applied.

## 3. Improved basic arithmetic circuits

Basic arithmetic circuits perform elementary arithmetic operations between two variables: addition, subtraction, multiplication and division, and are a necessity towards building a universal RPC-based computer.

As will be shortly seen, addition and multiplication can be performed without approximation, and with relatively simple circuits. On the other hand, division and subtraction, known so far, use approximate approaches which result in erroneous calculation. The question is whether the precision can be improved and at which cost. To preserve advantages of RPC, in the design of novel circuits one should take care of the following requirements:

1. Minimize calculation error over the whole state space of input parameters;
2. Minimize deviation from Poissonian process at the output;

3. Minimize quantity of hardware required to build the circuit.

Of course, these 3 requirements are generally pair-wise exclusive, thus generating a Mexican stand-off situation. Therefore, generating new and/or improved circuits for RPC is not trivial.

In the DC, addition and subtraction have exactly the same complexity in terms of execution time and number of gates required, while multiplication and division are substantially more complex. Division is a bit more complex than multiplication because it requires an iterative procedure over the other three elementary operations.

Strangely, in RPC (with unipolar RPTs) the simplest operations are multiplication and addition, while division and subtraction are substantially more complex. Subtraction is a bit more complex since, as will be shown, it can be made by a non-iterative use of division and multiplication.

### 3.1 Multiplication

Multiplication is the simplest operation in RPC: it can be done with a single AND gate, as shown in Fig. 2. Because pulses from RPTs $p_0$ and $p_1$ are independent, a probability to have two such pulses in the same time bin is just $p_0\, p_1$, therefore this circuit gives the exact result. Note that both input and output variables have span in the interval $[0,1]$, which means that this circuit is optimal in terms of dynamic range (no scaling needed).

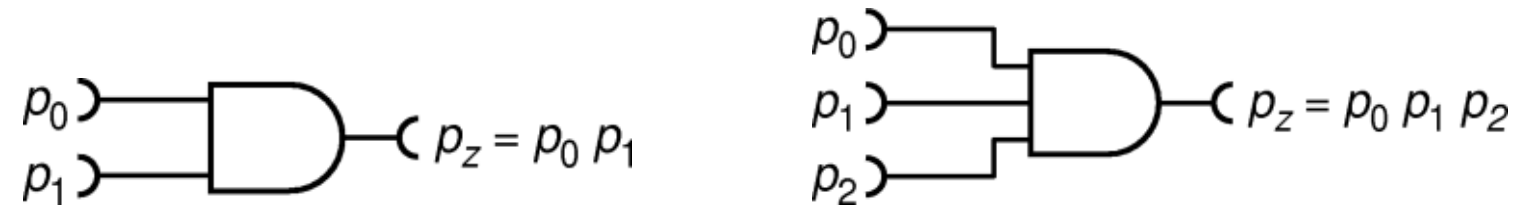

a) b)

**Fig. 2.** Circuits for exact multiplication of two numbers (a); and three numbers (b).

This principle can be upgraded to multiply three or more numbers *at the same time* with a minimum additional hardware requirement, just by adding the required number of inputs to the AND gate, as shown in Fig 2b.

This is illustration of one of the key differences between DC and RPC. While calculation in DC is sequential, i.e. multiplication of several numbers assumes first multiplying two of them, then multiplying the result with the third one etc. until numbers are exhausted, in RPC the multiplication is done simultaneously. Furthermore, there is no idle time during which no information is present at the output. The former is also known as a "streamline" computation. It is easy to imagine why both characteristics are favorable for survival.

### 3.2 Summation

Approximate addition of two RPTs can be done with a single OR gate, as shown in Fig. 3a. Any input pulse pass through OR gate, but if the two pulses coincide, only one pulse will be formed at the output. Following the discussion from the previous section, it is plain to see that operation done by the OR gate is:

$$p_z = p_0 + p_1 - p_0 p_1 \,. \tag{2}$$

It is interesting that $p_z \in [0,1]$, which can be seen immediately from equality:

$$p_z = 1 - (1 - p_0)(1 - p_1). \tag{3}$$

Thus, this circuit is dynamically optimal, albeit it does not compute addition but something else. For small $p_0$ and $p_1$ this circuit performs an approximate addition because then the multiplicative term in Eq. (2) may become negligible. This circuit can be generalized to perform a simultaneous approximate summation of 3 or more numbers by just adding inputs to the OR gate. For example, for RPTs with probabilities $p_0, p_1, p_2$ a 3/input OR gate would calculate:

$$p_z = 1 - (1 - p_0)(1 - p_1)(1 - p_2) = p_0 + p_1 + p_1 - p_0 p_1 - p_1 p_2 - p_2 p_0 + p_0 p_1 p_2 \quad (4)$$

which is again dynamically optimal and calculates approximate summation of the three RTPs.

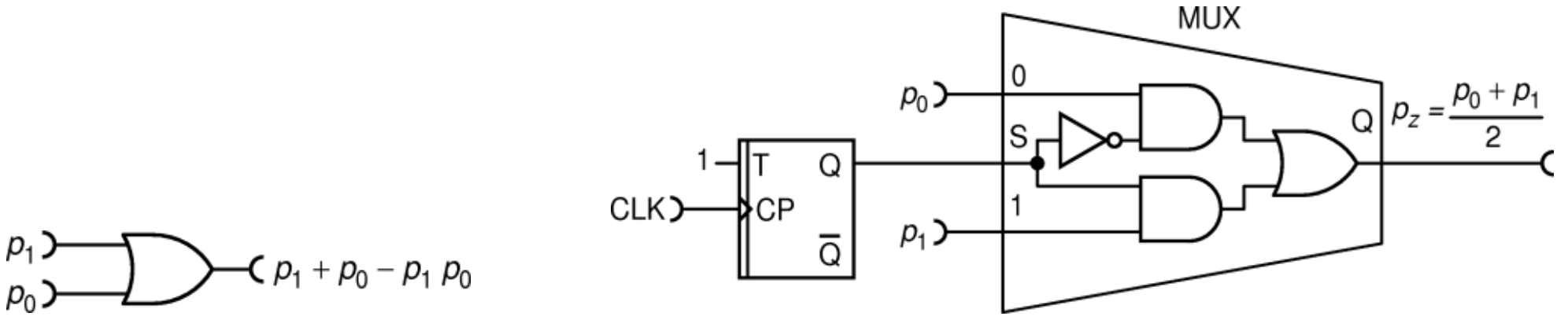

a) b)

**Fig. 3.** Addition circuits: approximate with OR gate (a); exact with a multiplexer (MUX) circuit (b).

Exact addition with a dynamically optimal scaling can be done via a multiplexer (MUX) circuit, shown in Fig. 3b. This circuit samples randomly, with equal probability, pulses from the two inputs. In previous art, thus, the resulting operation is:

$$p_z = \frac{p_0 + p_1}{2}. \quad (5)$$

and circuit is dynamically optimal. The extra factor 2 is not a caveat because it can be counted in a calculation. However, a caution has to be exercised because this kind of binary (2-input) addition is not associative:

$$\frac{1}{2}\left(\frac{1}{2}(p_0 + p_1) + p_2\right) \neq \frac{1}{2}\left(p_0 + \frac{1}{2}(p_1 + p_2)\right). \quad (6)$$

This circuit can be generalized to sum $n$ inputs by using one-out-of-$n$ random MUX circuit, but such a circuit is not trivial. In fact, it is not currently known how to make such a MUX, unless $n$ is a power of 2.

### 3.3 Division

Division is a difficult operation in RPC and only approximate methods are known. Let us imagine we want to calculate $p_z = p_0/p_1$. The idea is to start with a guessed value (probability) of $p_z$ (say 0.5), multiply it with $p_1$ (we do know how to multiply exactly!) and compare result to $p_0$. If the result is smaller than $p_0$ we enlarge $p_z$. On the contrary, we take a smaller $p_z$ until the right solution is reached. A circuit that does that, published in [10] is shown in Fig. 4a. Probably due to a lapsus, this circuit is wrong: it generates pulses of variable length and their abundance is far too small. Our correction of this circuit, shown in Fig. 4b, works correctly.

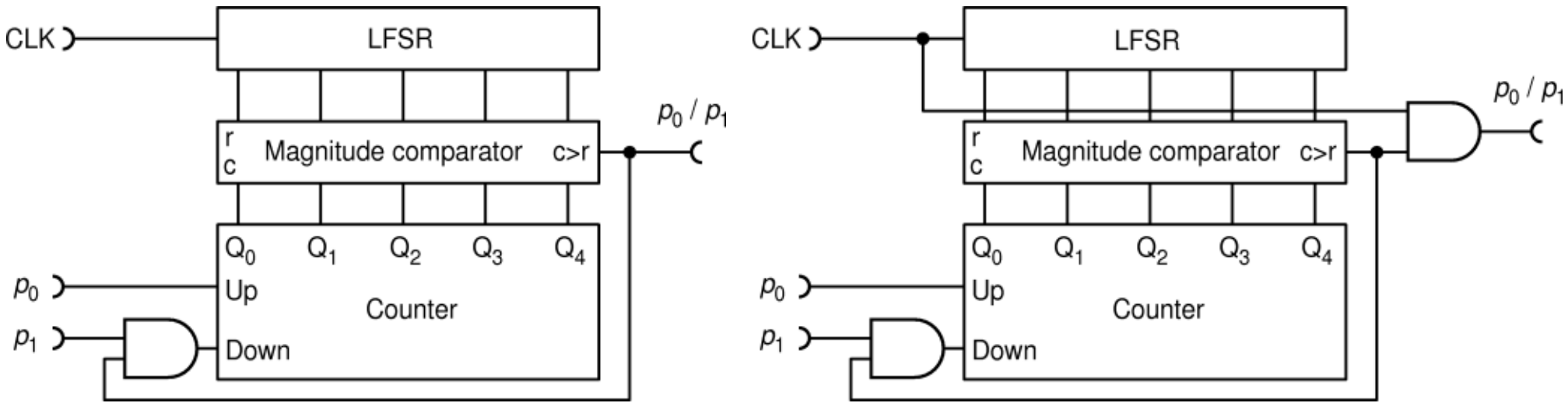

a) b)

**Fig. 4.** Feedback based division circuit: erroneous version (a); correct version (b).

To perform the above algorithm, one needs three distinct logic sub-circuits: an integer random number generator (here realized with a linear feedback shift register, LFSR), a magnitude comparator and a counter, all of the same length of $N$ bits. We note that in this work we always assume that neither can an $N$-bit counter cannot count over its maximum $(2^N - 1)$ nor below zero.

At this point one could ask whether counters exist in live neurons. It turns out that they do. In fact, a probability that a neuron will fire a pulse is proportional to the difference between number of excitatory and inhibitory pulses received and thus a neuron behaves as an Up/Down counter in which excitatory pulses increase, while inhibitory decrease its state! Literature is vague on how many pulses can a neuron count before going into saturation, but some works mention thousands [11], which could account for up to 10-12 bits.

The transfer function and errors for a certain space of input parameters $p_0$, $p_1$ and the counter bit of length $N$, for the divider in Fig. 4a are shown in Fig. 5.

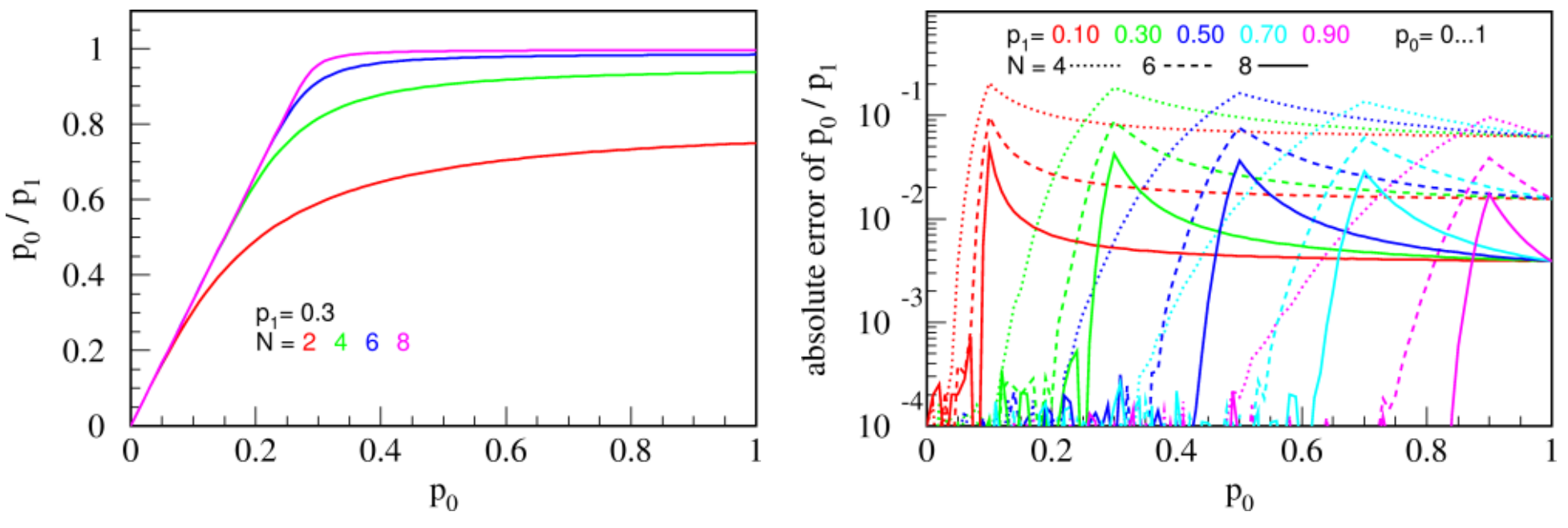

**Fig. 5.** The transfer function (a); and errors from the ideal division calculation (b) for the circuit shown in Fig. 4b.

When $p_1 > p_0$, the function levels off at the maximum possible value of 1 as shown in Fig. 5a. This behavior is accounted for in the graph of errors shown in Fig 5b. We see that error is largest when $p_1 \approx p_0$. The error becomes smaller with enlarging counter's number of bits $N$, but it remains rather large even for an 8-bit counter.

A problem with this circuit is that, due to the use of a LFSR which shifts by one bit with each CLK pulse, it tends to cause long sequences of pulses and long periods of absence of pulses at the output $p_z$, thus deforming the desired exponential p.d.f. This will have a deteriorating impact to precision of calculations in which output of this circuit is used as an input. One way to mitigate this problem would be to use as many different LFSRs as there are bits in the comparator, but that would greatly add to the complexity of the circuit, aggravate the problem of seeding LFSRs and still not eliminate possibility of correlations among output pulses due to pseudorandomness.

In order to avoid these pitfalls, we propose to use true randomness implemented via RFFs, as shown in Fig. 6. According to [5], a T-type RFF (TRFF) acts as an ordinary TFF with provision that its clock input (Cp) acts with probability of 0.5, randomly. Thus, if T input is held high, a TRFF generates an independent random bit upon each clock pulse.

Indeed, we find that this circuit performs better in terms of randomness of the output that circuit in Fig 4b, but its precision differs insignificantly form the circuit in Fig 4b.

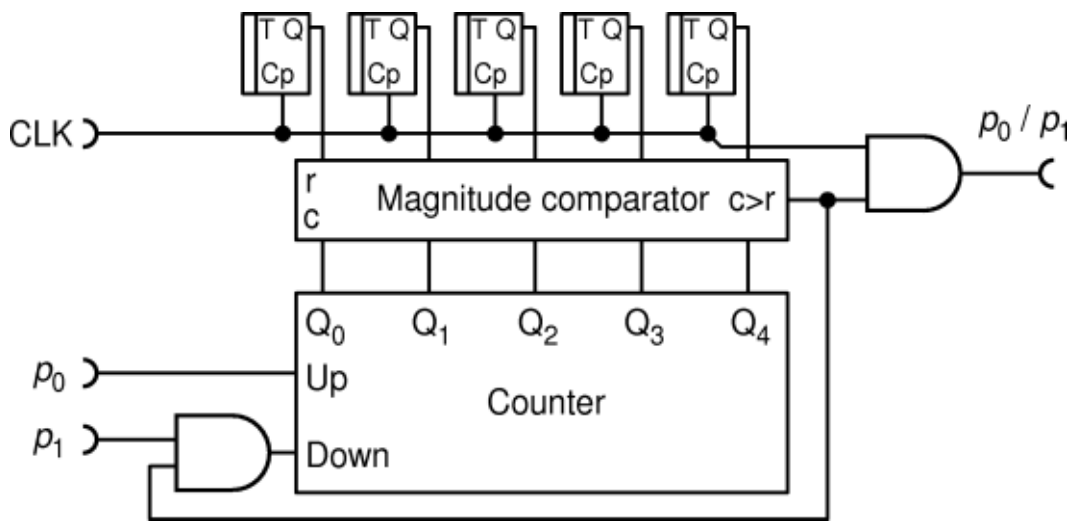

**Fig. 6.** T-type random flip flop (a); and its use in the improved division circuit (b). If not connected, input T is assumed being HIGH.

To improve greatly on precision and simplicity, we further propose circuit shown in Fig 7.

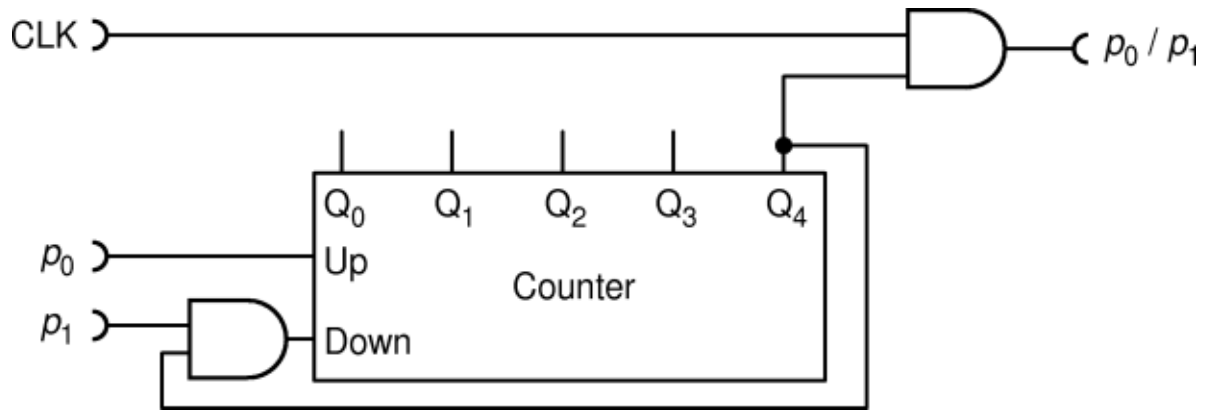

**Fig. 7.** A simple and precise division circuit.

This circuit is much more sensitive to errors and therefore keeps better precision at the output. Its transfer function and errors are shown in Fig. 8.

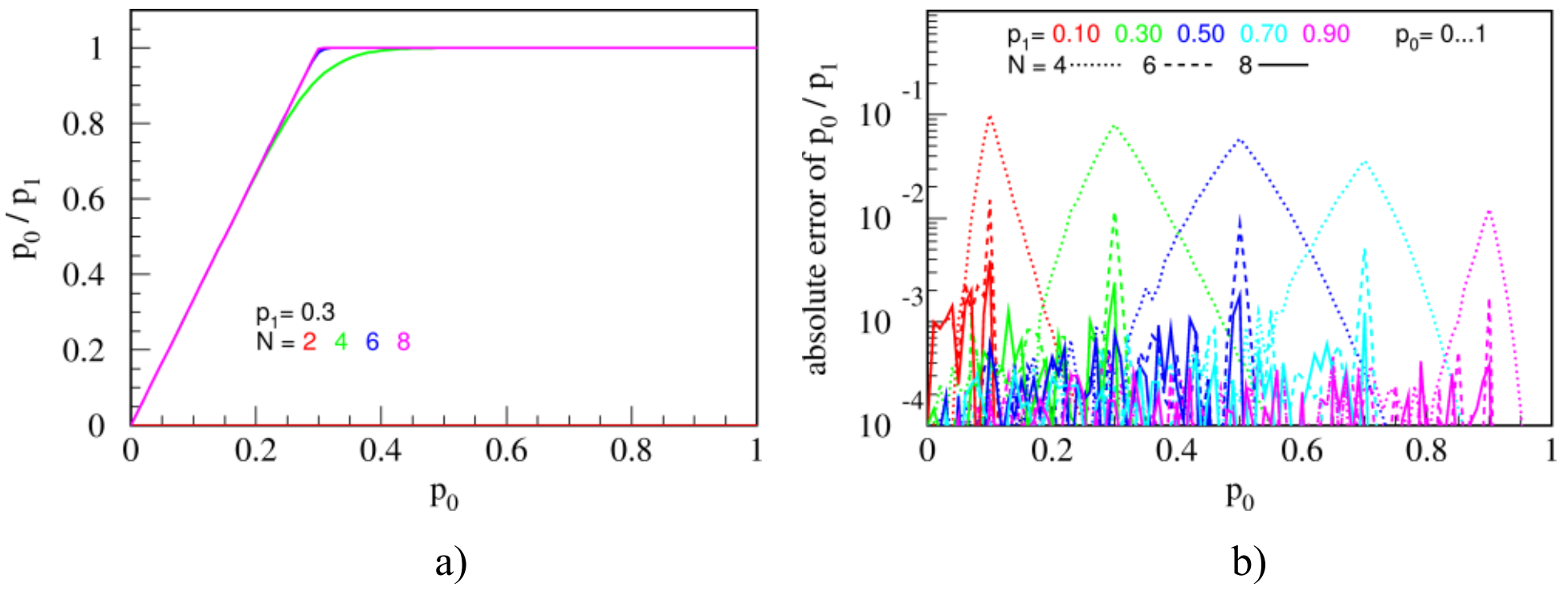

**Fig. 8.** The transfer function (a); and errors from the ideal division calculation (b) for the circuit shown in Fig. 7.

Because of its precision, circuit in Fig. 7 allows us to use a smaller counter, which further simplifies the division circuit. Note that it has no comparator nor sources of randomness, only the counter, but because of that it tends to generate very long sequences of pulses and absence of pulses, being the worst in terms of deviation from the exponential distribution.

### 3.4 Subtraction

Subtraction if the most complicated of all basic arithmetic functions in RPC paradigm. This can be understood by noticing the following expression:

$$p_1 - p_0 = \left(1 - \frac{p_0}{p_1}\right) p_1 \,. \tag{7}$$

Multiplication and subtraction from 1 can be easily done by logic circuits as shown in Fig 9a. By inserting the division circuit shown in Fig. 4b, one arrives to the subtraction circuit in Fig. 9b [10].

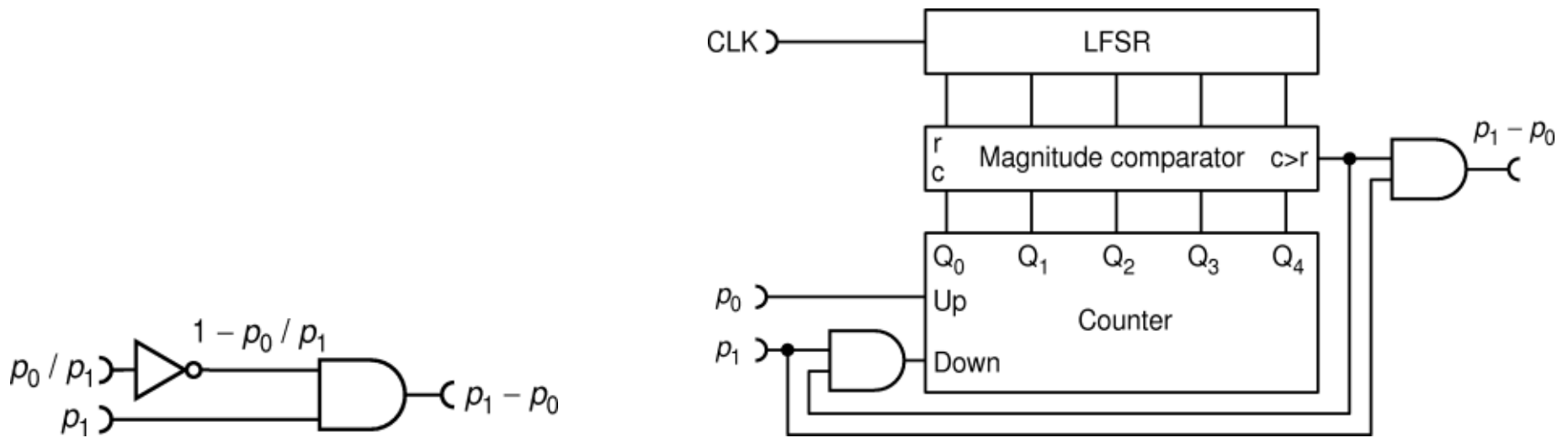

a) b)

**Fig. 9.** Subtraction via division principle (a); realization of a subtraction circuit utilizing a division circuit (b).

This circuit inherits properties of the division circuit. Again, improvement in randomness and hardware reduction, without a gain in precision, can be obtained by substituting LFSR with TRFF, as shown in Fig. 10.

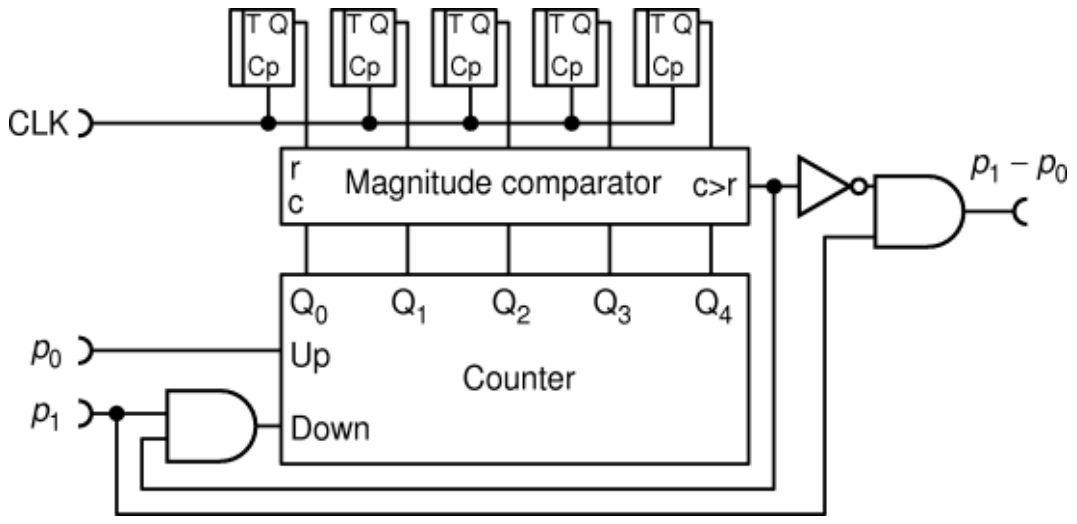

**Fig. 10.** A subtraction circuit with an improved output randomness.

Transfer function and errors of both circuits are shown in Fig. 11.

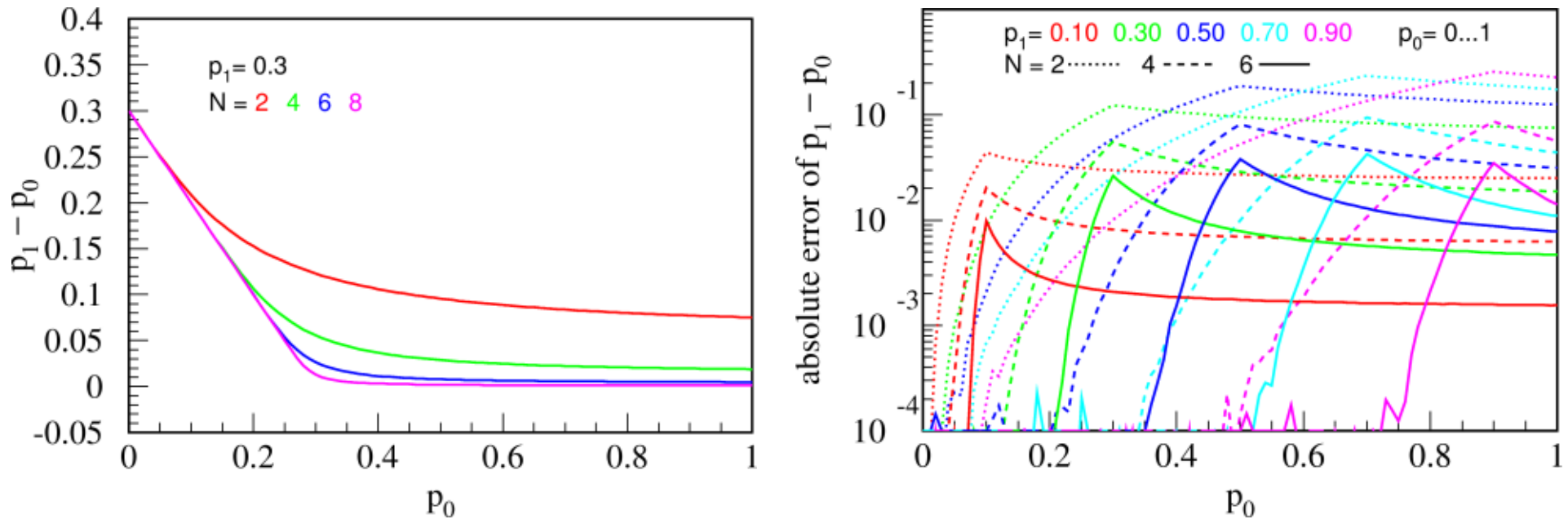

a) b)

**Fig. 11.** The transfer function (a); and errors from the ideal division calculation (b) for the circuit shown in Fig. 7.

We see that the transfer function in Fig. 11a contains two linear parts. A non-sharp joint of the two indicates computation error around $p_0 = p_1$, the problem inherited from the division circuit used herein and explained for graphs in Fig. 5 and circuit in Fig. 4a.

To significantly improve on precision, albeit with worsening of the randomness of the output, we propose the circuit in Fig. 12.

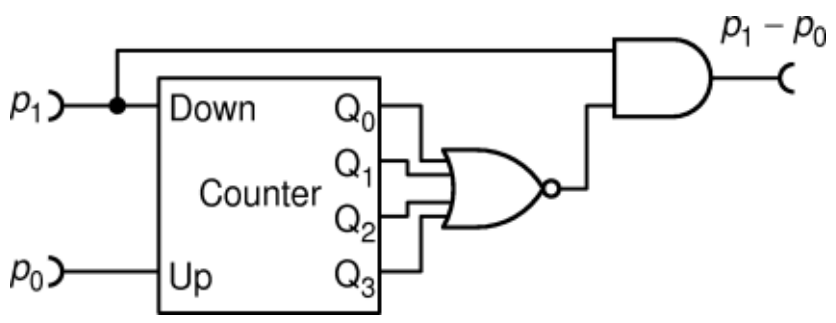

**Fig. 12.** A simple and precise subtraction circuit with counter of $N = 4$ bits.

The circuit functions in the following way. The counter counts pulses from $p_0$ and inhibits exactly as many pulses from $p_1$ by means of the output AND gate. A source of error is when counter does not have enough capacity to count pulses from $p_0$ before a pulse from $p_1$ arrives and goes into saturation. The chance of this happening is highest when $p_0 \approx p_1$ and can be lowered by taking a bigger counter.

Even though this subtractor is not derived from a divider, the transfer function and errors, shown in Fig. 13, show close resemblance to patterns seen in division circuit in Fig. 7.

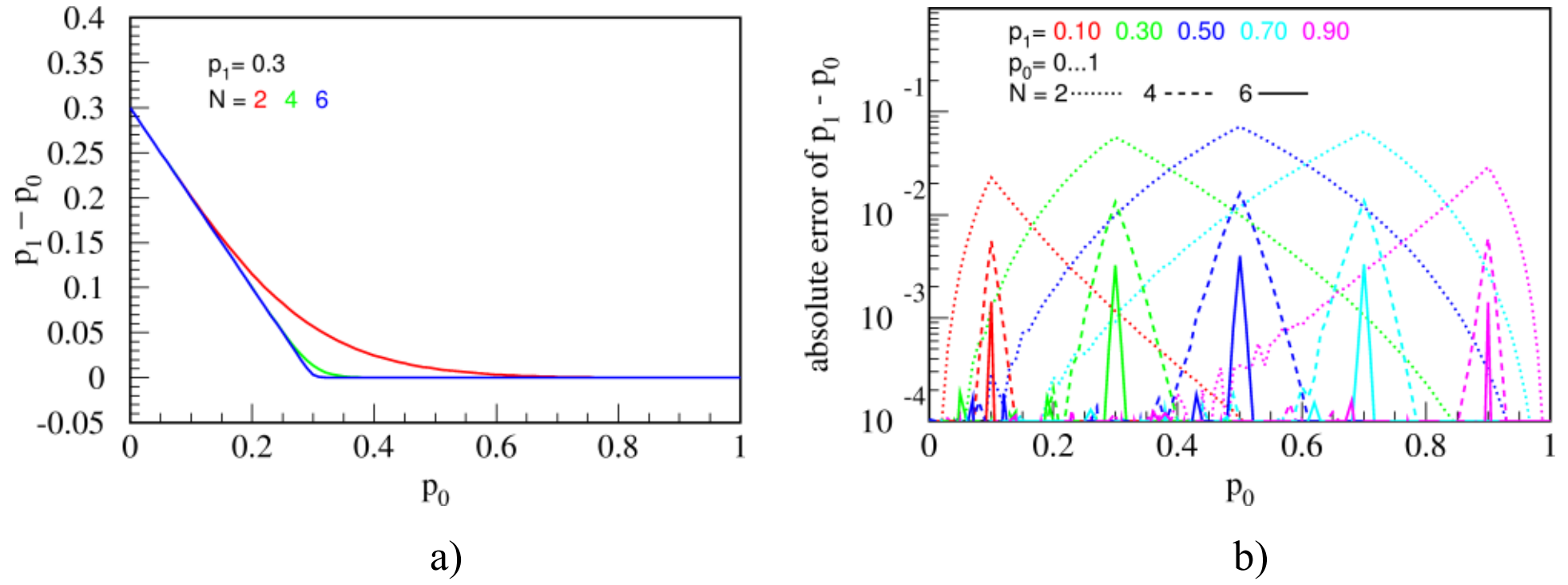

**Fig. 13.** The transfer function (a); and errors from the ideal division calculation (b) for the circuit shown in Fig. 12.

Again, we see there is a sharp edge between linear parts indicating small errors. In fact, explanation of transfer function and errors is quite similar of the division circuit in Fig. 7.

Perhaps division circuit in Fig. 7 and subtraction circuit in Fig. 12 are the most precise possible but with not-so-great output randomness. It may be that there is an inevitable trade-off between precision and randomness of the output. These are questions open for further investigation.

## 4. Magnitude comparator and flow control

A universal computing machine must allow for a flow control. This is usually done through a function that compares two numbers, such as $p_0 > p_1$, which returns TRUE or FALSE (logic 1 or 0 respectively), while in RPC it would return the RPT with $p = 1$ or $p = 0$ respectively.

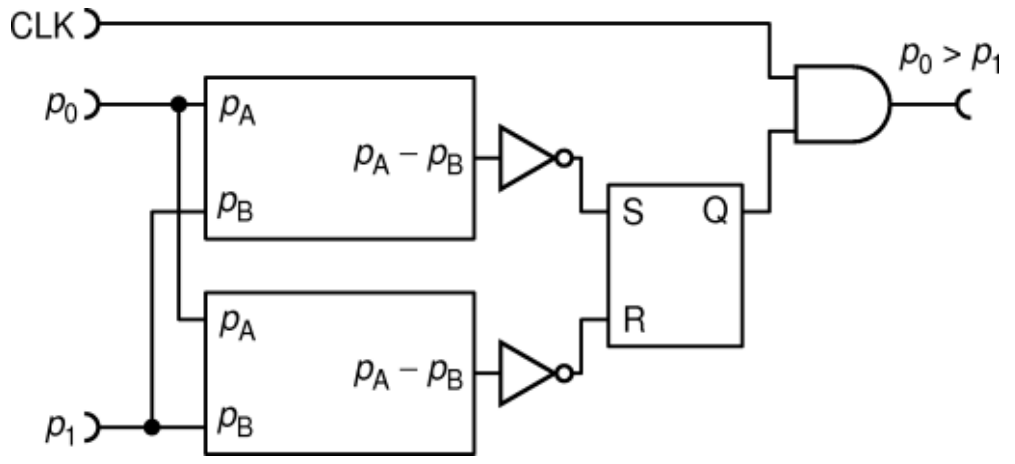

**Fig. 14.** Comparator circuit made of two identical subtracting circuits and a RS-type flip-flop.

First, we note that circuit in Fig. 12 yields 0 for all $p_0 > p_1$. But it is not very sharp: it won't switch to 1 as soon as $p_0 < p_1$, rather it would give a small value equal to $p_1 - p_0$. The solution is found in using two counter-connected subtractors and a RS-type flip-flop, as shown in Fig. 14. This circuit will switch to 1 sharply when $p_0 > p_1$and to 0 when $p_1 > p_0$. If a steady state instead of RPT is required/sufficient then the AND gate and CLK input can be omitted. Subtraction circuit of Fig 12 is ideal for use in the comparator because of its precision and simplicity, while its low randomness is of no consequence for the purpose.

## Acknowledgments

Funded by Croatian Ministry of Science Education and Sports, grants 533-19-14-0009 and KK.01.1.1.01.0001.